\documentclass[journal]{IEEEtran}
\usepackage[pdftex]{graphicx}
\usepackage{amssymb}
\usepackage{amsmath}
\usepackage{algorithmic}
\usepackage{array}
\usepackage[style=ieee, doi=false, isbn=false, url=true, eprint=false, maxnames=3]{biblatex}
\usepackage{xcolor}
\usepackage[caption=false,font=footnotesize]{subfig}
\usepackage{framed}  

\begin{document}
\title{GRAPPA-GANs for Parallel MRI Reconstruction}

\author{Nader~Tavaf, Amirsina Torfi, Kamil~Ugurbil,
        Pierre-Fran\c{c}ois~Van de Moortele
\thanks{Center for Magnetic Resonance Research (CMRR), University of Minnesota Twin Cities, Minneapolis,
MN, 55455 USA. E-mail: tavaf001@umn.edu.}
}

\markboth{Tavaf \MakeLowercase{\textit{et al.}}: GRAPPA-GANs for Parallel MR Image Reconstruction}%
{}

\maketitle

\begin{abstract}
\textit{k}-space undersampling is a standard technique to accelerate MR image acquisitions. Reconstruction techniques including GeneRalized Autocalibrating Partial Parallel Acquisition (GRAPPA) and its variants are utilized extensively in clinical and research settings.
A reconstruction model combining GRAPPA with a conditional generative adversarial network (GAN) was developed and tested on multi-coil human brain images from the fastMRI dataset. For various acceleration rates, GAN and GRAPPA reconstructions were compared in terms of peak signal-to-noise ratio (PSNR) and structural similarity (SSIM).
For an acceleration rate of R=4, PSNR improved from 33.88 using regularized GRAPPA to 37.65 using GAN. GAN consistently outperformed GRAPPA for various acceleration rates. 

\end{abstract}


\section{Introduction}

\IEEEPARstart{M}{agnetic} Resonance Image (MRI) is a prevalent non-invasive medical imaging technique with various clinical and research applications. A major advantage of MRI is its potentially high resolution; however, MRI generally requires lengthy acquisition times to achieve high resolution images. Undersampling the MR signal (obtained in frequency domain a.k.a. \textit{k}-space) is a method to accelerate such time-consuming acquisitions. Parallel imaging refers to the methods used for reconstructing MR images from undersampled \textit{k}-space signal. Generally, parallel image reconstruction techniques take advantage of the additional encoding information obtained using (ideally independent) elements of a receiver array and/or mathematical properties of the frequency domain signal to compensate for the loss of information due to the undersampling. Nevertheless, consequences of that information loss generally detract from the quality of the images reconstructed from undersampled \textit{k}-space.

The aim of improving the undersampled reconstructions can be pursued from multiple different angles. While an extensive review of all such research efforts is beyond the scope of this article, we still mention a few relevant works in each line of research to provide context for the current paper.
In terms of hardware, there has been significant effort in the MR research community to improve the sensors used to acquire the signal (radio-frequency coils) to reduce noise and noise correlation between different channels or to take advantage of additional receive channels (e.g. \cite{Keil2013_massively, Tavaf2020a, Shajan2016, Adriany2019}). There has been a wider variety of advancements in the post-processing front.
SENSE~\cite{Pruessmann1999} and GRAPPA~\cite{Griswold2002} are two of the primary methods for parallel MR image reconstruction. GRAPPA tries to estimate the missing \textit{k}-space signal but it inherently suffers from noise-amplification. Generally, the \textit{k}-space undersampling comes at the expense of aliasing in reconstruction. 
Several variations and extensions to SENSE and GRAPPA have been proposed which primarily rely on regularization to suppress noise-amplification. Compressed-sensing also relies on non-linear optimization of randomly undersampled \textit{k}-space data, assuming the data is compressible~\cite{Donoho2006}. Compressed sensing MRI generally utilizes total variation, wavelet/cosine transforms, or dictionary learning as sparse representations of the naturally compressible MR images. 

More recently, side effects of existing techniques (noise amplification, staircase artifacts of total variation, block artifacts of wavelets, relatively long reconstruction time of iterative optimization techniques, etc) and the advent of public MR image datasets have encouraged researchers to look into deep learning techniques which have often outperformed conventional regularization and/or optimization-based techniques in various applications, including variants of the undersampled image reconstruction problem (e.g. \cite{Liang2020,Torfi}). Among the promising literature, several works have used generative adversarial networks (GANs)~\cite{Goodfellow, Goodfellow2020} to reconstruct undersampled images. Yang et al.~\cite{Yang2018} proposed a GAN to address the aliasing artifact resulting from the sub-Nyquist sampling rate. Their proposed architecture used a pretrained network to extract an abstract feature representation from the reconstruction and enforce consistency with the target in that feature level. Murugesan et al.~\cite{Murugesan2019} and Emami et al.~\cite{Emami2020} used context dependent/attention-guided GAN which has a feedback loop back to the generator input providing information focusing on local deviations from tissue. Mardani et al.~\cite{Mardani2019} and Deora et al.~\cite{Deora} used residual skip connections inside each convolutional block of their generator. It is noteworthy that Mardani suggests the discriminator outputs can be used to focus on sensitive anatomies. Dar et al.~\cite{Dar2020} also used perceptual priors in their multi-contrast reconstruction GAN. The above mentioned studies using GANs have demonstrated enhanced performance compared to state of the art compressed sensing and other parallel imaging reconstruction techniques. However, one of the primary critiques of GAN-based reconstruction is the suggestion that GANs are prone to hallucination (see for example ~\cite{Mardani2019}). 

Here, we propose a novel method for reconstruction of undersampled/accelerated MRI images that combines GRAPPA and GAN to further improve the reconstruction quality by building on our proof-of-principle demonstration~\cite{Tavaf2021}. Our primary contributions include:
\begin{itemize}
    \item we propose a combination of GRAPPA and GAN, 
    \item in addition to the adversarial losses, we include data-consistency and perceptual feature level loss for artifact removal.
\end{itemize}

\section{Methods}
\begin{figure}
    \centering
    \includegraphics[width=3in]{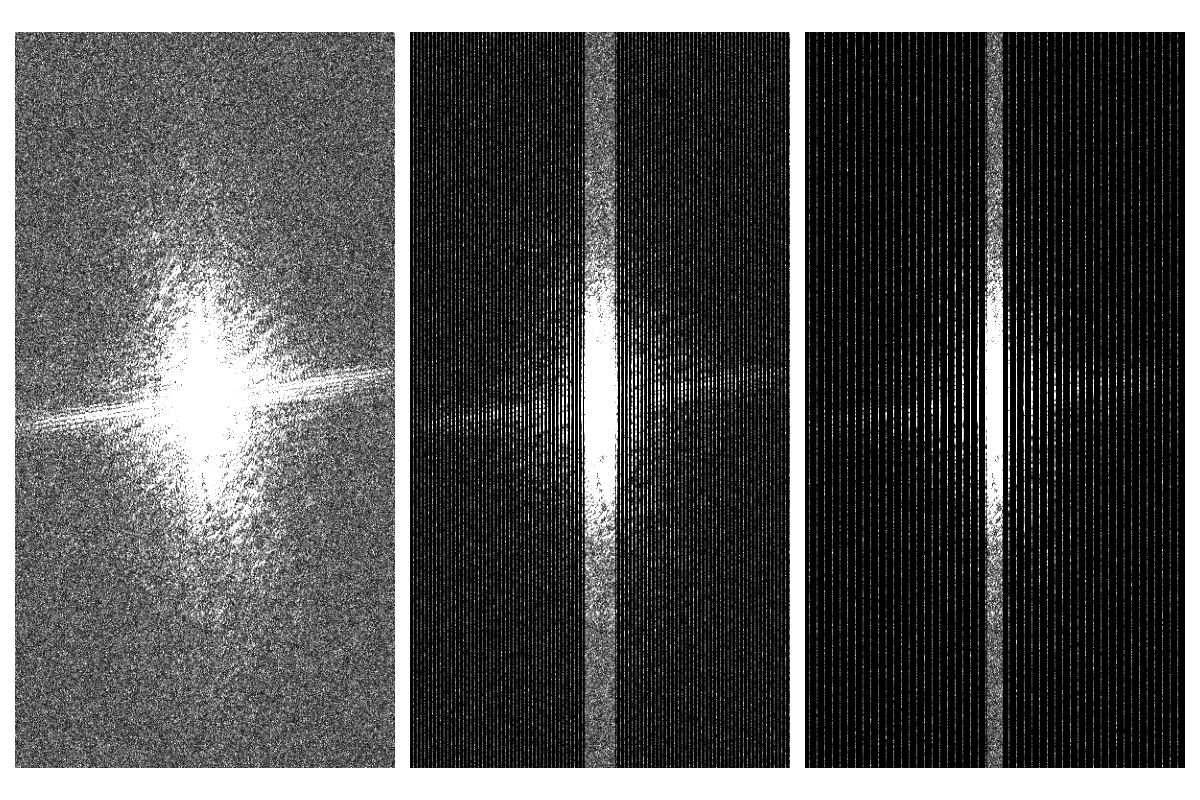}
    \caption{Equidistant \textit{k}-space undersampling with random position of the first \textit{k}-space line, keeping the central \textit{k}-space fully-sampled (ACS lines used for GRAPPA). From left to right: fully sampled, subsampled with R=4, subsampled with R=8.}
    \label{fig:kspace_undersampling}
\end{figure}

\begin{figure*}
    \centering
    \includegraphics[width=6in]{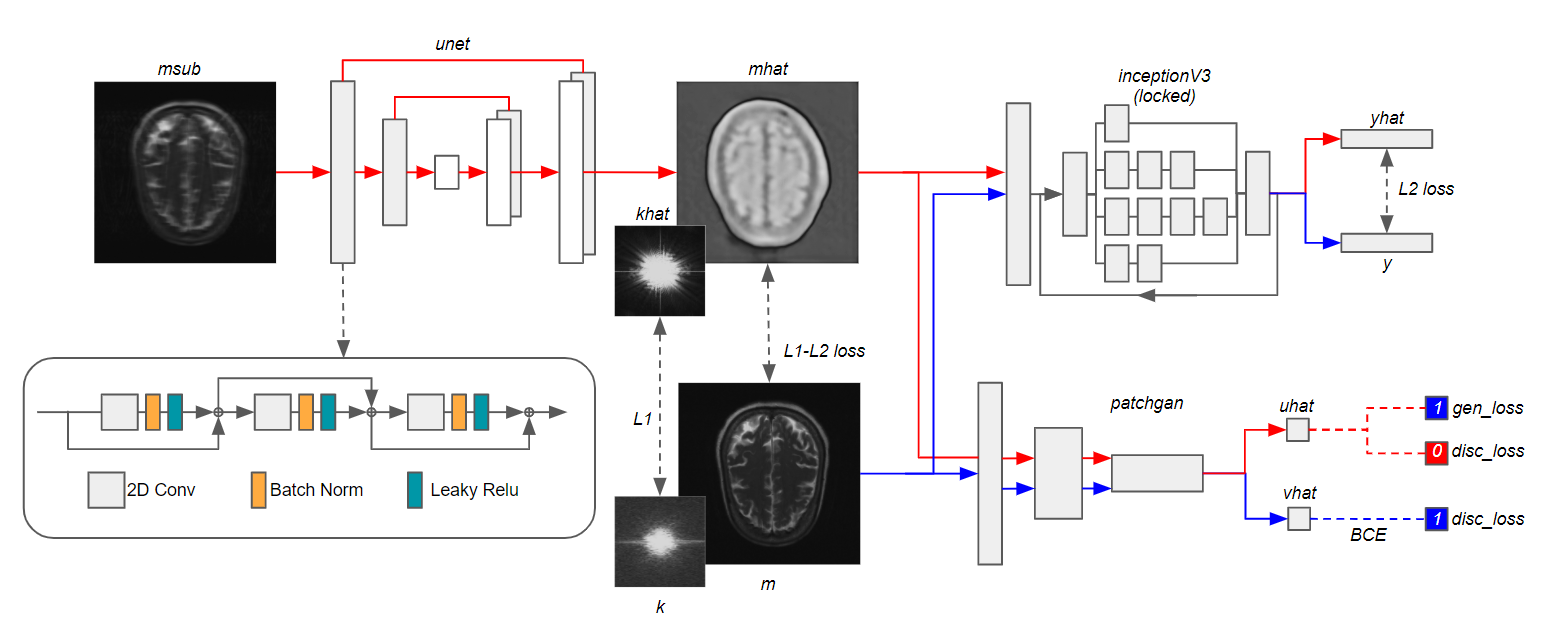}
    \caption{Symbolic network architecture. The UNET consisted of five levels, starting with 64 channels at the first layer. Kernel size used for 2D convolutions was 3x3 (in both R=4 and R=8 experiments, due to computational limitations). The InceptionV3 network was pretrained on ImageNet and used to extract and compare features from the generator output and target image. Each convolution block of the UNET consisted of three layers of convolution, batch normalization, leakyRelu interleaved with resnet-type skip connections.}
    \label{fig:architecture}
\end{figure*}

\subsection{Undersampling scheme}
The original data is fully sampled in \textit{k}-space, allowing for comparison of undersampled reconstructions with a fully-sampled ground truth reconstruction. Various undersampling schemes have been used in the literature, with uniform random subsampling, equidistant random subsampling, and Gaussian random subsampling being the primary schemes. Given that our dataset (discussed in more detail shortly) is composed of 2D axial slices, our analysis uses only 1D subsampling along the phase encoding direction. Here, we have used equidistant random subsampling while maintaining a fraction of the \textit{k}-space lines at the center of the \textit{k}-space fully-sampled, as is customary in the MRI literature and required for GRAPPA reconstruction. Equidistant random undersampling means that while the \textit{k}-space is subsampled equidistantly, the location of the first \textit{k}-space line is selected at random. For an acceleration rate (or subsampling ratio) of R=4, 8\% of \textit{k}-space lines were preserved at the center and for R=8, 4\% of the \textit{k}-space lines were preserved at the center. Figure~\ref{fig:kspace_undersampling} demonstrates the subsampling scheme in \textit{k}-space.

\subsection{Reconstruction method}
Details of GRAPPA implementations have been included in various publications~\cite{Griswold2002}. Briefly, GRAPPA uses linear shift-invariant convolutions in k-space. Convolutional kernels were learned from a fully sampled subset at the center of k-space (auto-calibration signal or ACS lines) constrained by a Tikhonov regularization term and then used to interpolate skipped k-space lines using multi-channel (receive array) raw data. We did a GRAPPA root-sum-of-squares reconstruction of the undersampled, multi channel\footnote{The term ``channel'' in the MRI context refers to the number of sensors or coils used in image acquisition, whereas in the deep learning context, it is used interchangeably with number of kernels or filters.} image prior to feeding it to the GAN.
In a generic GAN, a generator network ($G: m_\_\rightarrow \hat{m}$) competes with a discriminator ($D:\hat{m}\rightarrow (0, 1)$) in a min-max optimization problem, \\ 
$\displaystyle \min_{\theta_G}\max_{\theta_D}\mathcal{L}(\theta_D, \theta_G) = \mathrm{E}[\log D(\hat{m})] + \mathrm{E}[\log(1-D(G(m_{\_}))]$, \\where the generator learns the mapping from the GRAPPA reconstruction of the undersampled image, $m_\_$, to its prediction, $\hat{m}$, of the target, fully sampled image, $m$. Note that the GAN is learning in image domain (not the frequency domain). 

In essence, first, regularized GRAPPA is used to fill-in the missing \textit{k}-space lines. Then, 2D discrete fast Fourier transform is performed to reconstruct individual images of individual coils. A root-sum-of-squares (RSS) reconstruction, $m_\_$, of the individual magnitude images from individual coils is then used as the input to the generator. The generator learns to predict the ground-truth given this subsampled reconstruction while the discriminator learns to classify / distinguish between generator-reconstructed images and ground-truth images.

The GAN was composed of a generator (a UNET~\cite{Ronneberger2015}) and a discriminator (a convolutional neural network used as a binary classifier). 
The network architecture is depicted symbolically in Figure~\ref{fig:architecture}. The UNET consisted of an encoder and a decoder. The encoder was composed of blocks of batch normalization~\cite{Ioffe2015}, 2D convolution, and leakyReLu, interleaved by max pooling to down-sample the images. Each one of these blocks had three convolutional/activation layers with in-block (resnet type) skip connections passing the information derived at earlier layers to the features computed at later layers. The decoder was composed of similar normalization, convolution, leakyReLu blocks interleaved by transpose 2D convolutions for up-sampling. Skip connections were used to add high-level feature representations of the encoding path to elements of the decoding path. The original implementation in~\cite{Ronneberger2015} learns a prediction of the image, however, we included a skip connection from the input of the encoder to be added to the output of the decoder, so that the UNET is learning the residual (difference). Residual learning (compared to learning the full reconstruction task) proved to be a less challenging task, requiring less model complexity. Furthermore, the addition of the in-block skips noticeably improved performance results. Depth of the UNET was five levels, with the top level limited to 64 kernels at most (due to hardware limitations) and 3x3 convolutional kernels.

The discriminator was topped with a dense layer and sigmoid activation appropriate for the binary classification of images (classifying generator reconstructions versus ground truth) using binary cross entropy loss. In addition to the typical generator GAN loss (binary cross entropy of the discriminator judgment of generator output compared with ones, or $-\log[D(\hat{m})]$), the generator loss was conditioned on a weighted sum of L1 and L2 loss terms comparing generator output with target reconstruction, a data-consistency loss term comparing the output and ground truth in spatial frequency domain (k-space), and an inception loss, comparing the InceptionV3~\cite{Szegedy2016} feature representation of generator output and ground truth. Overall, this results in, \\
$\mathcal{L}(\theta_G) = \log(D(\hat{m})) + \lambda_1 L_1(\hat{m}, m) + \lambda_2 L_2 (\hat{m}, m) + \lambda_{DC} L_1(\mathcal{F}(\hat{m}), \mathcal{F}(m)) + \lambda_f L_2(\mathcal{I}(\hat{m}), \mathcal{I}(m) ) $\\
where $\mathcal{F}$ is the Fourier transform that maps the images to frequency domain, and $\mathcal{I}$ is the Inception network used to extract features. Note that the Inception network was pretrained on ImageNet~\cite{Russakovsky2014} and locked (no weight updates) during training. In other words, the InceptionV3 network was used only to calculate a perceptual loss~\cite{Ledig2016}, that is used to evaluate the performance of the generator (or to accentuate feature level irregularities of generator reconstruction), not as part of the generator's architecture, and need not be used in deployment. In the absence of the Inception feature loss, the L1-L2 loss would focus on pixel level similarity, which is useful in improving the performance metrics (discussed shortly), but leaves noticeable residual aliasing artifacts in the reconstruction. The focus on feature loss (at later epochs of training) helped resolve these residual aliasing artifacts. The addition of the frequency domain data consistency loss helped capture the higher spatial frequency details of the anatomy.

\subsection{Dataset}
The data used in this work were obtained from the NYU fastMRI Initiative database, with detailed descriptions of the datasets published previously in~\cite{Knoll2020b, Zbontar}. In the present study, we used multi-coil, multi-slice human brain images from the fastMRI dataset. As this dataset includes a variety of real-world acquisitions (with different MR scanners, protocols, artifacts, contrasts, radio-frequency coils, etc) and because variation in each of these factors (especially the number of coils) would cause significant variation in the results, we selected a subset of the dataset limited to images acquired with 16 receive coils\footnote{The choice of ``16" was because it was the largest subset of the dataset and it was appropriate for an acceleration factor of R=4, and still reasonable for R=8.}. This removed a parameter that would otherwise significantly affect variance in results and therefore, made result interpretation more straightforward. Other than number of coils, and ensuring no subject overlap between train/validation/test sets, no other constraint was imposed on the multi-coil human dataset. The original data were fully sampled. The accelerations (subsampling) were imposed as post-processing steps.

\subsection{Evaluation metrics}
Peak signal-to-noise ratio (PSNR) and structural similarity (SSIM) were used to assess the performance~\cite{Wang2004}. The reconstructions were compared with a ground truth, defined as root-sum-of-squares reconstruction of fully sampled k-space data from individual channels. PSNR was calculated as $-20\log10(RMSE/L)$ where RMSE is the root-mean-square error and L is the dynamic range. SSIM was calculated as $\frac{\left(2\mu_x\mu_y+c_1\right)\left(2\sigma_{xy}+c_2\right)}{\left(\mu_x^2+\mu_y^2+c_1\right)\left(\sigma_x^2+\sigma_y^2+c_2\right)}$ 
using an 11x11 Gaussian filter of width 1.5 and $c_1,c_2$ of 0.01, 0.03 respectively.

\subsection{Training and implementation details}
Individual loss terms were normalized to be on similar scales. Training started with a focus on L1 similarity, with $\lambda_1=120, \lambda_2=30, \lambda_{DC}=0, \lambda_f=0$. Midway through training (30 to 50 epochs), the weight balance of L1-L2 loss gradually changed to $\lambda_1=30, \lambda_2=120$. After 100 epochs, the focus shifted to feature loss and data consistency loss while maintaining the L1-L2 weights, with $\lambda_{DC}=30, \lambda_f=100$. 

The GAN was trained using 100 subjects (1600 axial slices) while the validation and test dataset each included an additional 100 subjects, without any subject overlap between the three subsets. An Adam optimizer~\cite{Kingma2015Adam:Optimization} with a customized learning rate schedule was used. Custom python scripts were used for GRAPPA and GAN implementations, with the GAN implemented using TensorFlow 2.2 / Keras. The network was trained for 200 epochs using one NVIDIA Tesla V100 GPU.

\section{Results}
\begin{figure}
    \centering
    \includegraphics[width=3in]{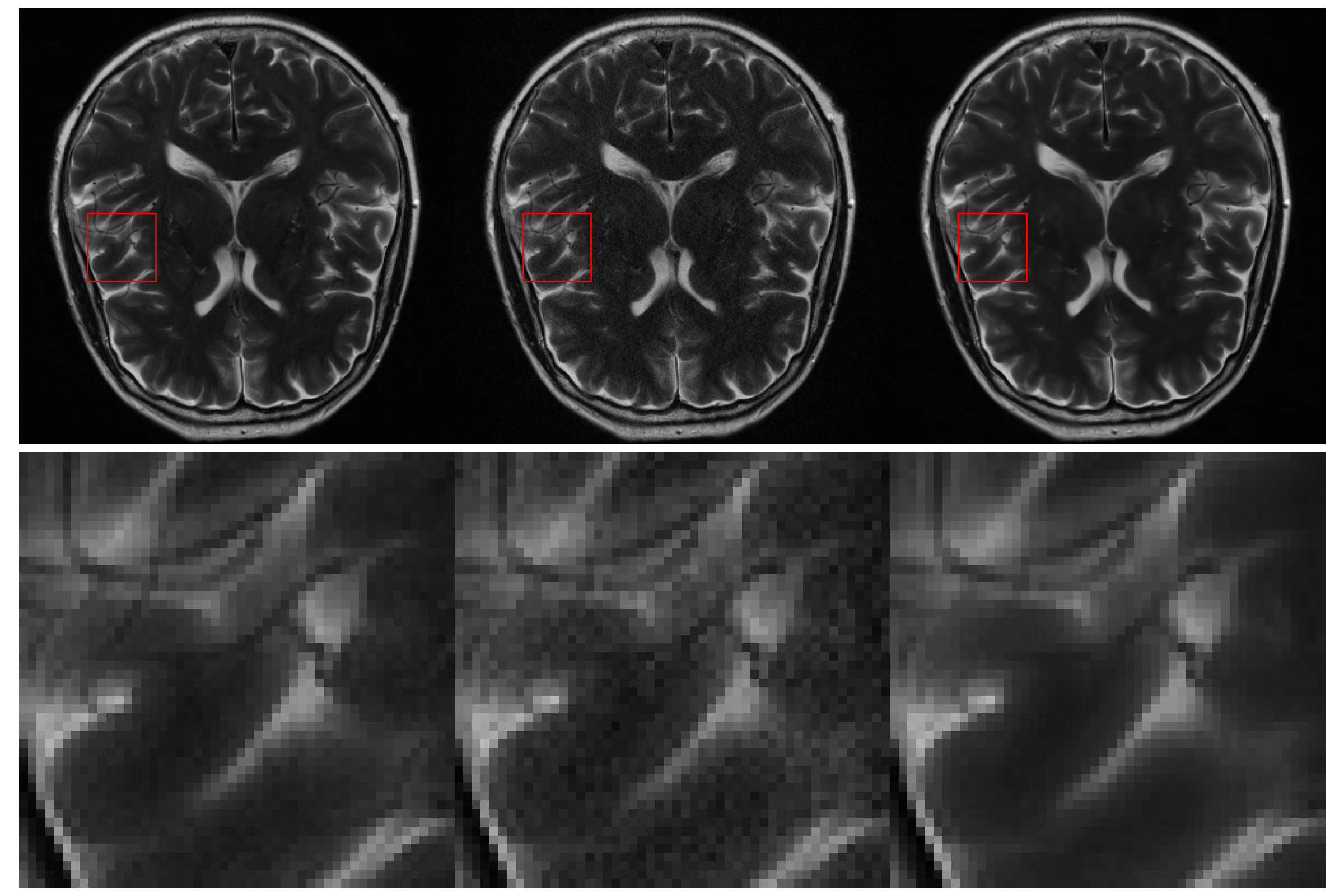}
    \caption{Comparing reconstruction quality at acceleration factor R=4. Left: ground truth (fully sampled root-sum-of-squares reconstruction); center: regularized GRAPPA reconstruction (uniform undersampling, 8\% ACS lines); right: GAN reconstruction.}
    \label{fig:R4_comparison}
\end{figure}
\begin{figure}
    \centering
    \includegraphics[width=3in]{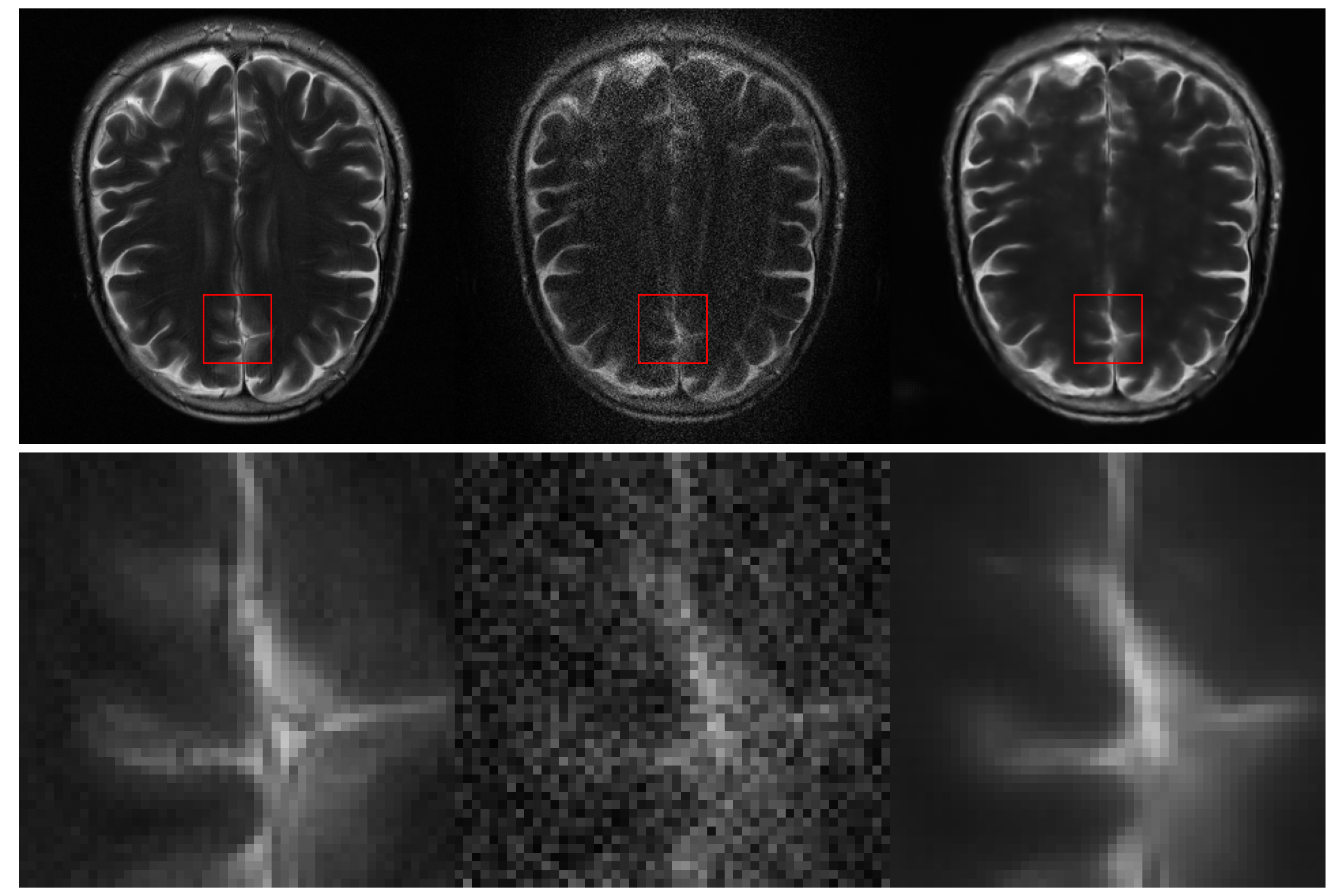}
    \caption{Comparing reconstruction quality at acceleration factor R=8. Left: ground truth (fully sampled root-sum-of-squares reconstruction); center: regularized GRAPPA reconstruction (uniform undersampling, 4\% ACS lines); right: GAN reconstruction.}
    \label{fig:R8_comparison}
\end{figure}

Figure~\ref{fig:R4_comparison} and Figure~\ref{fig:R8_comparison} present a qualitative comparison between reconstructions using regularized GRAPPA and GPGAN. As presented in Table 1, with an acceleration factor of R=4, regularized GRAPPA resulted in PSNR=33.88dB and SSIM=0.84. The GAN improved the results to PSNR=37.65dB and SSIM=0.93. 
The average root-mean-square error reduced from 0.021 to 0.013 for R=4 and from 0.075 to 0.033 for R=8, using GRAPPA and GAN, respectively.  
The increase in SSIM is due to reduced standard deviation ($\sigma_x$) of the GAN reconstruction, suggesting a higher statistical signal-to-noise ratio (SNR $\propto$ mean(signal) / std(noise)) using GAN.


\begin{table}[t!]
    \centering
    \caption{Comparing average performance results for different 1D acceleration factors (R) with regularized GRAPPA and GAN.}
    \begin{tabular}{|c|c|c|c|c|}
        \hline
         &  \multicolumn{2}{c|}{R=4} & \multicolumn{2}{c|}{R=8}\\
         \cline{2-5}
         & PSNR & SSIM & PSNR & SSIM \\
         \hline
         GRAPPA & 33.88 & 0.84 & 22.45 & 0.51 \\
         \hline
         GAN & 37.65 & 0.93 & 29.64 & 0.84\\
         \hline
    \end{tabular}
    \label{tab:results_compare}
\end{table}

\section{Discussion}
\begin{figure}
    \centering
    \subfloat[]{\includegraphics[width=0.24\textwidth]{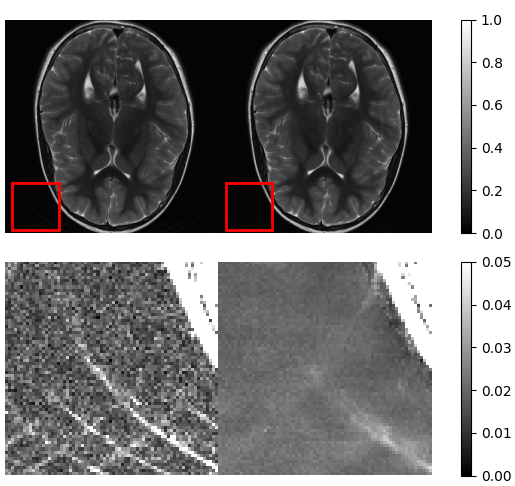}\label{fig_first_case}}
    \hfill
    \subfloat[]{\includegraphics[width=0.24\textwidth]{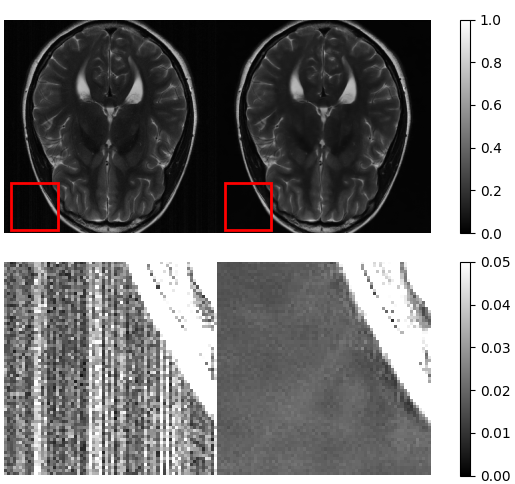}\label{fig_second_case}}
    \caption{Denoising and artifact suppression using the proposed GAN. In both (a) and (b), the left subfigures are the ground truth and the right subfigures are the GAN reconstructions. The lower row are the zoomed-in and rescaled detail view of the respective red boxes.}
    \label{fig:denoise_artifact}
\end{figure}

While the primary purpose of the proposed technique is reconstruction of sub-sampled k-space (i.e. addressing the aliasing artifact), the fully sampled dataset was contaminated with other common real-world artifacts (Gibbs artifacts, motion artifacts, etc.) which were often mitigated in the final GAN reconstruction. Figure~\ref{fig:denoise_artifact} illustrates artifact suppression. Moreover, the GAN reconstruction was effective in denoising reconstructions and improving the average statistical signal-to-noise ratio of the images. Incorporating GRAPPA into the data-driven reconstruction pipeline improves the structural fidelity of the reconstructed images, making sure that no significant structures are added or deleted in the final result (although some details are inevitably lost due to undersampling). 

While the dataset included acquisitions using various numbers of receiver channels (from 8 to 23 receive channels), in order to prevent high variance in accelerated reconstructions due to variance in receiver channel count, we used only a subset of the dataset including only acquisitions with exactly 16 receive channels. Nevertheless, an acceleration factor of R=8 using only 16 receive channels results in significant noise in the GRAPPA reconstruction. By comparison, the GAN reconstructions are noticeably less noisy even with R=8 acceleration.

Building on previous works~\cite{Rahmaninejad2020, Buffinton2020, Habibian2019, Habibian2021}, various elements of the generator loss function ensure different aspects of the reconstruction fidelity. The perceptual prior imposed using the inception network is aimed to achieve feature level consistency. This ensures that prominent features of the reconstruction follow the same distribution as the target dataset. While this helps eliminate the residual aliasing artifacts, it also captures and tries to replicate other real-world artifacts of the target dataset. The latter is mitigated by the data consistency loss term.


In future, we would like to build upon this work by integrating a GAN with a compressed-sensing solution of the image reconstruction problem.

\section{Conclusion}
A generative adversarial network was used to improve the quality of accelerated MR image reconstruction using regularized GRAPPA. The results demonstrate significant reduction in root-mean-square error of accelerated reconstruction compared with the fully sampled ground truth.

\section{Acknowledgements}
The authors acknowledge funding from NIH U01 EB025144, P30 NS076408 and P41 EB027061 grants.

\AtNextBibliography{\small}
\printbibliography

\end{document}